\documentclass[conference,a4paper]{IEEEtran}

\usepackage{graphicx,amsmath,epsfig,amssymb,amsthm,latexsym,mathrsfs,epsf,enumerate,bm,color}
\usepackage{bbm}
\newsavebox{\fminibox}
\newlength{\fminilength}

  \def\+{^\dagger}
\def\nequiv{\not\kern-.05em\equiv}
\def\egal{\kern-.5em=\kern-.5em}        % Moins d'espace autour de "="
\def\propt{\kern-.2em\propto\kern-.2em} % Idem
\def\argmax{\mathop{\mathrm{arg\,max}}} % Mieux que \def\argmax{\arg\max}
\def\argmin{\mathop{\mathrm{arg\,min}}} % car l'indice est reparti
\def\intdouble{\int\kern-0.3em\int}
\def\inttriple{\int\kern-0.3em\int\kern-0.3em\int}

\def\rond#1{\overset{\kern-0.33em~_\circ}{#1}}
\def\rondit[#1]#2{\overset{\kern#1~_\circ}{#2}}

\usepackage{algorithmic}
\usepackage{array}
\usepackage{epstopdf}

\newtheorem{Theorem}{Theorem}
\newtheorem*{Theorem*}{Theorem}

\newtheorem*{Claim*}{Claim}

\newtheorem{CounterExample*}{$\overline{\hbox{\bf Example}}$}
\newtheorem{Definition}{Definition}
\newtheorem*{Definition*}{Definition}

\newtheorem{Example*}{Example}

\newtheorem{Intuition*}{Intuition}
\newtheorem{Joke*}{Joke}

\newtheorem{Lemma*}{Lemma}

\newtheorem{Note*}{Note}
\newtheorem{Open problem}{Open problem}
\newtheorem{Proposition}{Proposition}

\newtheorem{Question*}{Question}
\newtheorem{Remark*}{Remark}

\newcommand{\exc}{{\rm  E}}

\newcommand{\uY}{\underline{Y}}
\newcommand{\uX}{\underline{X}}
\newcommand{\uy}{\underline{y}}
\newcommand{\ux}{\underline{x}}

\newcommand{\uW}{\underline{W}}
\newcommand{\uZ}{\underline{Z}}
\newcommand{\uz}{\underline{z}}
\newcommand{\ueta}{\underline{\eta}}
\newcommand{\uzm}{\underline{0}}

\newcommand{\ubsQ}{\underline{q}}

\newcommand{\mcD}{\mathcal{D}}
\newcommand{\mcT}{\mathcal{T}}
\newcommand{\mcN}{\mathcal{N}}
\newcommand{\mcV}{\mathcal{V}}
\newcommand{\mcE}{\mathcal{E}}
\newcommand{\mcF}{\mathcal{F}}
\newcommand{\mcG}{\mathcal{G}}
\newcommand{\mcY}{\mathcal{Y}}

\newcommand{\mcZ}{\mathcal{Z}}

\newcommand{\mbbR}{\mathbb{R}}

\newcommand{\bH}{{\bf H}}
\newcommand{\bD}{{\bf D}}
\newcommand{\bS}{{\bf S}}

\newcommand{\bC}{{\bf C}}

\newcommand{\bOmega}{\mathbf{\Omega}}

\newcommand{\bSigma}{\mathbf{\Sigma}}
\newcommand{\bSigmaT}{\mathbf{\widetilde{\Sigma}}}

\newcommand{\sumY}{\sum_{\uy \in \mcY}}

\begin{document}
%\title{A Greedy Approuch to Approximate Covariance Matrix of Renewable Generations with First Order Markov Chain by Minimizing Different Regret}
\sloppy

\title{Latent Tree Approximation in Linear Model}


%\author{\IEEEauthorblockN{-\\-\\-\\-\\-}
%\IEEEauthorblockA{-}}

\author{\IEEEauthorblockN{Navid Tafaghodi Khajavi}
\IEEEauthorblockA{Dept. of Electrical  Engineering, %\\
University of Hawaii, Honolulu, HI 96822\\
Email: navidt@hawaii$.$edu}}

%\and
%\IEEEauthorblockN{Toshihisa Tanaka}
%\IEEEauthorblockA{Dept. of Electrical and Electronic Engineering\\
%Tokyo University of Agriculture and Technology\\
%Tokyo, Japan\\
%Email: tanakat@cc$.$tuat$.$ac$.$jp}

\maketitle

\begin{abstract}
%\boldmath
We consider the problem of learning underlying tree structure from noisy, mixed data obtained from a linear model.
To achieve this, we use the expectation maximization algorithm combined with Chow-Liu minimum spanning tree algorithm. This algorithm is sub-optimal, but has low complexity and is applicable to model selection problems through any linear model.

\end{abstract}


% \fi
\IEEEpeerreviewmaketitle

%\vspace{-0.2cm}
\section{Introduction}

This paper considers learning in graphical models where we approximate a graphical model with a tree.  The tree approximations are made as it is much simpler performing inference and estimation on trees rather than graphs that have cycles or loops.  An example is applying the Gaussian belief propagation (GBP) algorithm \cite{BP}.  When you have statistical knowledge of the graphical structure (i.e. correlation matrix), then you can use the Kullback Liebler  (KL) divergence measure and find the optimal tree using the  Chow-Liu algorithm which finds the minimal spanning tree (MST) \cite{ChowLiu}.  
There are other methods discussed in literature where instead of minimizing KL divergence, a detection framework is proposed in order to quantify the tree approximation quality \cite{MQ16ArXiv} and \cite{ITA2016}. Moreover, a similar method with a more generalized model is proposed in \cite{Allerton16}.
This paper considers the case when you do not have direct statistical information about the graph.  Here we assume you have an underdetermined and noisy linear model that gives access to samples of
data and we develop an algorithm using the EM algorithm and the Chow Liu algorithm to come up with a tree approximation by minimizing the KL divergence.

%Learning latent tree from noisy mixed observation is an interesting problem with lots of applications.
%Most of the distributed algorithm runs on trees or speed up when the underlying graph has a tree structure.
%One of the examples is Gaussian belief propagation (GBP) algorithm \cite{BP} which converges to the maximum likelihood solution if the underlying graph has a tree structure.
%In contrast, for graph with large number of loops, loopy GBP may not converge.
%As an example, loopy GBP has been used to do distributed state estimation for smart grid in \cite{ISIT2013}.
%To over come the convergence issue, in \cite{ISIT2013} the distribution of solar data approximated with a first order Markov chain distribution based on \cite{burg} in order to decrease the number of loops in loopy GBP.

Given a set of data from a jointly Gaussian random vector, the conventional tree approximation algorithms' goal is to approximate a tree structured distribution for the underlying distribution.
A first approach, is the well-known minimum spanning tree (MST) problem solved initially by Chow-Liu \cite{ChowLiu} which finds the covariance matrix coefficients using the covariance selection method \cite{dempster}. This algorithm is originally proposed to find the optimal tree structure that connects all the graph nodes and minimizes the KL divergence \cite{cover} between the original and the approximate distributions.
In contrast, in many applications such as the smart grid application \cite{ISIT2013}, we want to approximate structure (tree or chain) for a sub-graph of system graphical representation from the observation without the knowledge of the latent variable covariance matrix while considering the effects of the whole system, i.e. we want to learn a tree structure for the latent vector covariance matrix in an underdetermined model.

In this paper, we investigate the problem of approximating a tree structured covariance matrix of the latent variable in an underdetermined linear model by constructing an iterative algorithm based on the expectation maximization (EM) algorithm.
We also suggest to learn the maximum number of the EM algorithm iterations to avoid overfitting. 
Simulation results are performed on solar irradiation data from the island of Oahu, Hawaii \cite{NREL}.
As a result we can see the performance improvement using the proposed EM MST algorithm.

The rest of this paper is organized as follows. The tree approximation problem as well as the linear mixing model are stated in section \ref{sec:ProState} while the proposed EM MST solution is presented in section \ref{sec:TreeApproxALG}.
Section \ref{sec:simulation} presents some simulation on real spatial solar irradiation data and discusses the goodness of proposed solution.
Finally, Section \ref{sec:conclusion} summarizes results of this paper as well as discussion on future works.
\\
{\it Notation:}
Upper case and lower case letters denote random variables and their realizations, respectively;
underlined letters stand for vectors; boldface upper case letters denote matrices;
$\left( \cdot \right)^{T}$ and $\exc \left( \cdot \right)$ stand
for transposition and expectation.

%%%%%%%%%%%%%%%%%%%%%%%%%%%%%%%%%%%%%%%%%%%%%%%%%%%%%%%%%%%%%%%%%%%%%%
%%%%%%%%%%%%%%%%%%%%%%%%%%%%%%%%%%%%%%%%%%%%%%%%%%%%%%%%%%%%%%%%%%%%%%

\section{Problem statement}
\label{sec:ProState}
We want to approximate a multivariate distribution by the product of lower order component distributions \cite{lewis_approx}.
For the purpose of tree approximation, the maximum order of these lower order distributions is two, i.e. no more than pairs of variables.
Let $\uX \sim \mcN (\uzm , \bSigma)$ has a zero-mean Gaussian distribution where $\bSigma$ is the positive definite covariance matrix and $\uX \in \mbbR^p$ has the graph representation $\mcG=(\mcV, \mcE)$ where sets $\mcV$ and $\mcE$ are the set of all vertices and edges of the graph representing of $\uX$.
Let $\uX_{\mcT} \sim \mcN (\uzm , \bSigmaT)$ has the graph representation $\mcG_{\mcT}=(\mcV, \mcE_{\mcT})$ where $\mcE_{\mcT} \subseteq \mcE$ is a set of edges that represents a tree structure and $\bSigmaT$ is the positive definite covariance matrix of the tree structure.
Distribution of $\uX_{\mcT}$ (the product of lower order component distributions) is defined as \cite{ChowLiu}
\vspace{-0.2cm}
\begin{equation}
\label{eq:prod_approx}
f_{\uX_{\mcT}}(\ux_{\mcT}) = \prod_{(u,v) \in \mcE_{\mcT}} \frac{f_{\uX^u,\uX^v}(\ux^u , \ux^v) }{f_{\uX^u}(\ux^u) f_{\uX^v}(\ux^v)} \prod_{o \in \mcV} f_{\uX^o}(\ux^o)
\end{equation}
where $\uX^u$ and $\ux^u$are $u$th element of random vector, $\uX$, and its realization vector, $\ux$, respectively.
Let $\uY \in \mbbR^m$ be the observation vector where $m < p$.
Consider the problem of approximating a tree structured distribution of the latent vector, $\uX$, (similar to \eqref{eq:prod_approx}) in the following linear model\footnote{Without loss of generality, we may assume that $\uX$ and $\uY$ are zero-mean.}:
\begin{equation}
\label{eq:linear_model}
\uY = \bH \uX + \uW
\end{equation}
where $\bH$ is an $ m \times p$ characteristic matrix with rank $m$ and $\uW$ is a Gaussian zero-mean random vector with positive definite covariance matrix $\bD$.
Note that, in the linear model \eqref{eq:linear_model}, the latent vector, $\uX$, and noisy measurements, $\uW$, are independent and Gaussian, thus the observation vector, $\uY$, has Gaussian distribution.
We assume that the covariance matrix of noisy measurements, $\bD$, and the characteristic matrix $\bH$ are known completely but we may only have {\it partial prior information}\footnote{For instance, very noisy or outdated data directly obtained about $\uX$.}
% We will discuss in section \ref{sec:simulation} the possibility of achieving partial prior information about the latent vector, $\uX$.
 about the latent vector $\uX$.

%We seek to approximate the underling distribution of the latent vector $\uX$ such that it has a tree structure representation graph. In other words, instead of estimating the distribution of the original latent vector $\uX$, we seek to approximate the distribution of the tree structured latent vector $\uX_{\mcT}$ which obeys the product rule \eqref{eq:prod_approx} in the linear model \eqref{eq:linear_model}.

We seek to approximate the underlying distribution of the latent vector $\uX$ such that it has a tree structure representation graph. In other words, instead of estimating the distribution of the original latent vector $\uX$ and then approximate its tree structure, we seek to directly estimate the distribution of the tree structured latent vector $\uX_{\mcT}$ using a set of observed data obtained from the linear model \eqref{eq:linear_model}.

\section{Tree approximation algorithms}
\label{sec:TreeApproxALG}

In the sequel, we first consider the case where the complete knowledge of the distribution of the original latent vector $\uX$, i.e. the covariance matrix $\Sigma$ of the the zero-mean Gaussian distribution of $\uX$, is known. The optimal tree approximation in this case can be computed using the well-known Chow-Liu tree approximation algorithm \cite{ChowLiu}. 
This algorithm can also be applied to the estimated covariance matrix from the set of data observed directly form the latent vector $\uX$.
Then, we investigate the case where there is no information about the covariance of the latent variable.
In this case the tree approximation algorithm should be performed using the observation data obtained based on the model \eqref{eq:linear_model}. 
Since the linear model \eqref{eq:linear_model} is under-determined, i.e. $m < p$, there are infinitely many solutions.
We will show that by applying the expectation maximization (EM) algorithm, one can approximate a tree structured covariance matrix for the latent vector in this scenario.
The problem with this approach is that since the EM algorithm converging to local optima and there are infinitely many solutions, the EM algorithm may converge to a wrong local optima.
To deal with this problem, we assumed {\it partial prior information} is available to help the EM algorithm to converge to the desired optima.

\subsection{Chow-Liu Minimum Spanning Tree for Gaussian distributions}
Chow-Liu MST method \cite{ChowLiu} initially proposed for approximating the joint distribution of discrete variables by product of lower order distributions similar to \eqref{eq:prod_approx} which involves no more than the pair of variables and it can easily generalized for approximating the joint distribution of Gaussian variables.
\begin{Definition}
Let $\mcT^\star$ denote the set of all positive definite covariance matrices that their graphical models have tree structures and obey the product rule given in \eqref{eq:prod_approx}, i.e. set of all possible $\bSigmaT$'s that obey \eqref{eq:prod_approx}. \hfill \IEEEQED
\end{Definition}
In other words, for all $\bSigmaT \in \mcT^\star$, the corresponding zero-mean Gaussian distribution have the same marginals over the graph $\mcG_{\mcT}$ as the zero-mean Gaussian distribution with covariance matrix $\bSigma$. 
Note $\mcT^\star$ has finite cardinality of $O(p^{p-1})$. The KL divergence is proposed in \cite{ChowLiu} to quantify the distance between any distribution and its tree structure approximation. It is shown in \cite{ChowLiu} that the optimal solution for this problem can be found efficiently using greedy algorithms \cite{kruskal}.
\begin{Proposition}
The approximated tree structure for a zero-mean Gaussian distribution, $f_{\bSigmaT}(\ux)$ where $\bSigmaT \in \mcT^\star$, follows \eqref{eq:prod_approx} and thus the KL divergence can be simplified as follows:
\begin{equation*}
\label{eq:KLchowliu_sim}
\mcD ( f_{\bSigma}(\ux) || f_{\bSigmaT}(\ux) ) = - \frac{1}{2} \textnormal{log } (|\bSigma\bSigmaT^{-1}|). 
\end{equation*}
\proof Derivation is based on the properties of \eqref{eq:prod_approx} and the KL divergence definition for Gaussian distributions \cite{dempster}. \hfill \IEEEQED
\end{Proposition}

\subsubsection{Chow-Liu MST for a given distribution}
Let $f_{\bSigma}(\ux) =  \mcN (\uzm , \bSigma)$ be given. Then the Chow-Liu MST distribution is a zero-mean Gaussian distribution with covariance matrix, $\bSigmaT^\star \in \mcT^\star$, that minimize the KL divergence as follows:
\begin{equation}
\label{eq:chowliuopt_givenDist}
\bSigmaT^\star = \argmin_{\bSigmaT \in \mcT^\star} \; \mcD ( f_{\bSigma}(\ux) || f_{\bSigmaT}(\ux) ).
\end{equation}
where $\mathcal{D}^\star \triangleq - \frac{1}{2} \textnormal{log } (|\bSigma\bSigmaT^{\star -1}|)$ is the minimum and quantifies the distance between the given distribution and its optimal tree approximation.
Given the knowledge of $\bSigma$, Chow-Liu algorithm can efficiently compute the optimal solution, $\bSigmaT^\star$. 
Note that, finding the optimal solution $\bSigmaT^\star$ requires knowledge of the covariance matrix, $\bSigma$.
\\
{\bf Assumption:} In the rest of this paper, we assume that exact knowledge of the covariance matrix, $\bSigma$, is {\it not available}. We may only have some noisy or mixed observations. The main goal of this paper is to approximate the tree structured covariance matrix, $\bSigmaT$, using those noisy or mixed observations.

\begin{Definition}
Let $\mcT$ denote the set of all positive definite covariance matrices that have a tree structure. \hfill \IEEEQED
\end{Definition}
Note that, $\mcT^\star \subset \mcT$
%. Set $\mcT$ is a highly non-convex set on the cone of all positive definite covariance matrices. Also,
and its cardinality is infinite. In general, without exact knowledge of covariance matrix, $\bSigma$, it is not possible to find the optimal solution in \eqref{eq:chowliuopt_givenDist}  over the set $\mcT$.

% this section is after \end doc!

\subsection{EM Algorithm to Approximate MST}
Consider the linear model presented in \eqref{eq:linear_model}.  Let $\mcY :  \{ \uy_1,\ldots,\uy_R\}$ be a set of $R$ iid observation drawn from this model. Unlike previously mentioned models, one cannot uniquely compute an estimate of the covariance matrix, $\bSigma$, since this model is underdetermined.
In this section, instead of minimizing an optimization problem similar to \eqref{eq:chowliuopt_givenDist}, we suggest to minimize the KL divergence between the empirical distribution of observation obtained from \eqref{eq:linear_model}, i.e. set $\mcY$, and distribution of random vector $\uY_\mcT$, where the latent vector $\uX$ in \eqref{eq:linear_model} is replaced with the tree structured one, $\uX_\mcT$.

The empirical distribution of observation, $f_{\bS_{\uY}}(\uy) \sim \mcN (\uzm , \bS_{\uY})$  is a zero-mean Gaussian distribution where $\bS_{\uY} = \frac{1}{R}\sum_{r=1}^{R} (\uy_r-\bar{\uy})(\uy_r-\bar{\uy})^T$ where $\bar{\uy} = \frac{1}{R}\sum_{r=1}^{R} \uy_r$. 
The random vector $\uY_\mcT$ distribution is also zero-mean Gaussian distribution with covariance matrix $\bH \bSigmaT \bH^T + \bD$, i.e. $f_{\bSigmaT}(\uy) \sim \mcN (\uzm , \bH \bSigmaT \bH^T + \bD)$.
We minimize the KL divergence between these two distributions, in order to estimate the distribution of tree structured latent variable, $\uX_\mcT$. The optimization problem is as follows:
\begin{equation}
\label{eq:EMopt_KL}
{\bf \widetilde{\Sigma}}^* = \argmin_{\bSigmaT \in \mathcal{T} } \; \mathcal{D} \left( f_{\bS_{\uY}}(\uy) || f_{\bSigmaT}(\uy) \right). 
\end{equation}
Unlike previous cases, the Chow-Liu MST algorithm cannot find the optimal solution over set $\mcT$.
Basically, there are infinitely many solutions for the aforementioned optimization.
In what follows, we find a reasonable solution by implementing a heuristic algorithm which is a combination of the EM algorithm with the Chow-Liu MST algorithm.

The minimization problem in \eqref{eq:EMopt_KL} is identical to the following optimization problem which maximizes the average log-likelihood function where expectation is over the empirical distribution of observation, $f_{\bS_{\uY}}(\uy)$\footnote{Entropy of random variable $\uY$ with distribution $f_{\bS_{\uY}}(\uy)$ is eliminated here, since it does not depend on the optimization parameter, $\bSigmaT$.}:
\begin{equation*}
\bSigmaT^* = \argmax_{\bSigmaT \in\mathcal{T}} \exc_{\bS_{\uY}} \left[ \textnormal{log} \left( f_{\bSigmaT}(\uY) \right) \right].
\end{equation*}
Given the data set $\mcY$, we can write the average log-likelihood maximization as follows:
\begin{equation*}
\bSigmaT^* = \argmax_{\bSigmaT \in\mathcal{T}}\sumY \left[ \textnormal{log} \left( f_{\bSigmaT}(\uy) \right) \right].
\end{equation*}
Applying Jensen's inequality \cite{cover}, we have: 
\begin{align}\notag
\sumY \left[ \textnormal{log} \left( f_{ \bSigmaT }(\uy) \right) \right] &  = \sumY \left[ \textnormal{log} \left( \int_{\uX} f_{ \bSigmaT }(\ux,\uy) d\ux \right) \right] \\\notag
& = \sumY \left[ \textnormal{log} \left( \int_{\uX} g(\ux|\uy) \frac{f_{ \bSigmaT }(\ux,\uy)}{g(\ux|\uy)}  d\ux \right) \right] \\\notag
& \geq  \sumY \left[ \int_{\uX} g(\ux|\uy) \textnormal{log} \left( \frac{f_{ \bSigmaT }(\ux,\uy)}{g(\ux|\uy)}  d\ux \right) \right] \\
&  = \mcF \left( g(\ux|\uy) , \bSigmaT \right)
\label{eq:JensonBound}
\end{align}
where $f_{ \bSigmaT }(\ux,\uy)$ is the joint distribution between random vectors $\uX$ and $\uY$ and obviously depends on the approximation covariance matrix, $\bSigmaT$. 
Also, $g(\ux|\uy)$ is an auxiliary conditional distribution and $\mcF(.,.)$ is a function of the data set, $\mcY$, the auxiliary distribution, $g(\ux|\uy)$ and the approximation covariance matrix, $\bSigmaT$.

\begin{Definition}
Let $\mathcal{G}$ denote the set of all conditional Gaussian distributions, i.e. all possible $g(\ux|\uy)$ distributions. \hfill \IEEEQED
\end{Definition}
Using the EM algorithm, we can iteratively tighten the bound in \eqref{eq:JensonBound} by iteratively solving for $g(\ux|\uy)$ given $\bSigmaT$ and vice versa.
EM algorithm has two step at each iteration, the E-step and the M-step. The E-step of the $(l+1)$-th iteration of the EM algorithm is:
\\
{\bf E-Step:} computing $g(\ux|\uy) \in \mathcal{G}$ for all $\uy \in \mcY$:
\begin{align*}
g^{l+1}(\ux|\uy)  & = \argmax_{g(\ux|\uy) \in \mathcal{G}} \; \mcF ( g(\ux|\uy) , \bSigmaT^{l} ) \\
& = \argmin_{g(\ux|\uy) \in \mathcal{G}} \; \sumY \mathcal{D}( g(\ux|\uy) || f_{ \bSigmaT^l }(\ux | \uy) )\\
& = f_{ \bSigmaT^l }(\ux | \uy), \quad \forall \uy \in \mcY,
\end{align*}
where the tree structured covariance matrix $\bSigmaT^{l} \in \mcT$ is the $l$-th step solution of the EM algorithm.
The solution to the E-step is $g^{l+1}(\ux|\uy) = f_{ \bSigmaT^l }(\ux | \uy)$ for all $\uy \in \mcY$.
The M-step of the $(l+1)$-th iteration of the EM algorithm is:
\\
{\bf M-Step:} tightening the bound in \eqref{eq:JensonBound} by maximizing the lower bound for all $\bSigmaT \in \mcT$:
\begin{align*}
\bSigmaT^{l+1} & = \argmax_{\bSigmaT \in \mcT}\; \mcF \left( g^{l+1}(\ux|\uy) , \bSigmaT \right) \\
& = \argmax_{\bSigmaT \in \mcT} \;\sumY \left[ \int_{\uX} g^{l+1}(\ux|\uy) \textnormal{log} \left( f_{ \bSigmaT }(\ux , \uy) \right) d\ux \right]
\end{align*}
Plug in the solution of the E-Step into the M-Step optimization problem, we get:
\begin{align}
\bSigmaT^{l+1} & = \argmax_{\bSigmaT \in \mathcal{T}} \sumY \left[ \int_{\uX} f_{ \bSigmaT^l }(\ux | \uy) \; \textnormal{log} \left( f_{ \bSigmaT }(\ux , \uy) \right) d\ux \right]
\label{eq:EM_Lth_optT}
\end{align}
Optimization problem \eqref{eq:EM_Lth_optT} have to be solved at each iteration of the EM algorithm.

\begin{Theorem}
\label{Thm:thm1}
Solution to the $(l+1)$-th iteration of the EM algorithm (optimization problem \eqref{eq:EM_Lth_optT}) is given by:
\begin{equation*}
\bSigmaT^{l+1} = \texttt{chow-liu}(\bOmega^l)
\end{equation*}
where 
$\bOmega^l = \frac{1}{|\mcY|} \sumY  [\exc_{\bSigmaT^{l}} (\uX\uX^T | \uY=\uy )]$
where the expectation is over the conditional distribution, $ f_{ \bSigmaT^l }(\ux | \uy)$. For the linear model in \eqref{eq:linear_model}, we compute $\bOmega^l$ as
\begin{align*}
\bOmega^l & = \bC^{l} + \bC^{l} \bH^T \bD^{-1} \;\left(\frac{1}{|\mcY|} \sumY \uy \uy^T \right)\; \bD^{-1} \bH \bC^{l}
\end{align*}
where %% MERHAV %%
$\bC^{l} = \left( (\bSigmaT^{l})^{-1} + \bH^T \bD^{-1} \bH \right)^{-1}.$
\footnote{Proof is eliminated due to space limitation.}
\hfill \IEEEQED
\end{Theorem}

Having the solution of optimization problem \eqref{eq:EM_Lth_optT} in hand, the EM based algorithm is as follow:
\\
\hrule
{\bf EM MST Approximation Algorithm}
\hrule
\begin{itemize}
		\item{Initialization Step [$l=0$]:} 
				\begin{itemize}
					\item $\bOmega^{0} = \bSigma_{0}$
				\end{itemize}
		\item{Continue updating [$(l+1)$-th Step]:} 
				\begin{itemize}
					\item $\bSigmaT^{l+1} = \texttt{chow-liu}(\bOmega^l)$
					\item $\bOmega^{l+1} = \frac{1}{|\mcY|} \sumY  [\exc_{\bSigmaT^{l}} (\uX\uX^T | \uY=\uy )]$
				\end{itemize}
		\item Stopping criterion: $\mathcal{D} ( f_{\bSigmaT^{l}}(\ux) || f_{\bSigmaT^{l+1}}(\ux) ) < \epsilon$ \footnote{$\epsilon$ is some small number and have been set to avoid over-fitting.}
	  \item Output (approximated tree structure covariance matrix): $\bSigmaT_{EM}^* = \bSigmaT^{l+1}$ for some $l$ satisfying the stopping criterion
\end{itemize}
\hrule	
\vspace{0.5cm}
{\bf Remark:} The symmetric, positive definite matrix, $\bSigma_0$, is the initialization step in the EM MST approximation algorithm. 
One can choose any symmetric, positive definite matrix as an initialization step for this algorithm. 
However, EM based Algorithms converge to local minimum. Thus, having some prior knowledge will help the convergence of the algorithm.
We will discuss how to pick the matrix $\bSigma_0$ in section \ref{sec:simulation}.
The covariance matrix, $\bSigmaT_t^*$, is the approximated tree structure solution and is an upper bound for the optimal solution of \eqref{eq:EMopt_KL}, i.e., $\mathcal{D}  ( f_{\bS_{\uY}}(\uy) || f_{\bSigmaT_{EM}^*}(\uy) ) \geq \mathcal{D}  ( f_{\bS_{\uY}}(\uy) || f_{\bSigmaT^*}(\uy) )$.
\\
{\bf Remark:} The EM MST algorithm solution converges to one of the infinitely many solutions for the optimization problem in \eqref{eq:EMopt_KL}. 
But, the ultimate goal is to approximate a tree structured covariance matrix for the latent variable $\uX$ such that minimizes \eqref{eq:chowliuopt_givenDist}.
Since the distribution of $\uX$ is not available, we suggest to minimize \eqref{eq:EMopt_KL} which is a heuristic optimization problem based on the available set of observation from the underdetermined system.
Running the EM algorithm to find the exact solution for \eqref{eq:EMopt_KL} results in finding an overfitted solution for \eqref{eq:chowliuopt_givenDist}. To avoid overfitting, we need to learn the maximum number of iterations for the EM MST algorithm for each particular problem.

\section{Simulation Results and Discussion}
\label{sec:simulation}

In this section, we perform some simulations on real solar irradiation data obtain from \cite{NREL} to validate our results.
We take a year long data for $17$ horizontal sensors of the Oahu solar measurement grid sites \cite{APSIPA2014} from $9:00$ to $9:30$ AM to compute two normalized spatial covariance matrices\footnote{See \cite{APSIPA2014} for information about data and preprocessing methods involving computation of spatial covariances}.
We use data of Jan., 1st, 2011 to compute a covariance matrix, $\bSigma$, for testing the algorithm.
To run the algorithm we need some {\it partial prior information}.
In general, we initialized EM MST algorithm with different priors. For solar irradiation problem we could use a forecasting data or very noisy and outdated data or even without any irradiation data and by only knowing the geographical position of sensors, we can come up with an approximation for the covariance matrix only base on the location and use it as {\it partial prior information}.
Here, we use data from Apr., 1st, 2010 to Nov., 30th, 2010 to compute another covariance matrix, $\bSigma_0$, which can be used as prior to the EM MST algorithm.
In the linear model \eqref{eq:linear_model}, we assume vector $\uW$ is the vector of iid, white Gaussian noise.
We generate set $\mcY$ with $100$ samples with high signal-to-noise-ratio ($20$ dB) comparing to priors.

\begin{figure}[ht]
\begin{center}
\includegraphics[width=0.87\linewidth]{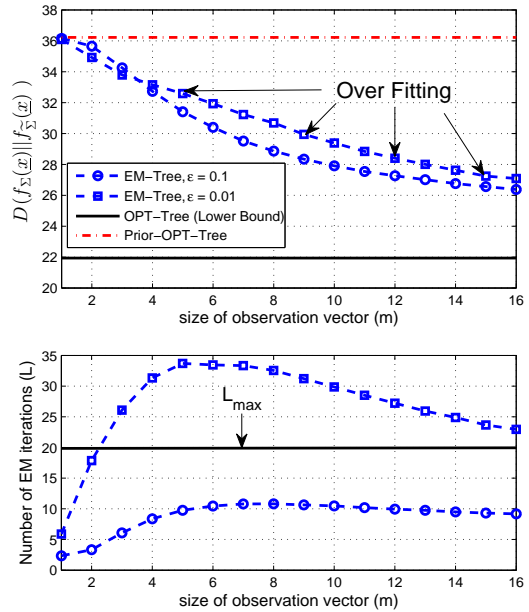}
\caption{{\bf Above:} The KL divergence between the approximated tree and the original graph vs. $m$ averaged over $1000$ randomly generated characteristic matrices, $\bH$, for the EM MST approximated tree with two different stopping criteria, $\epsilon=0.1$ and $\epsilon=0.01$, the optimal Chow-liu tree (lower bound) and the optimal tree for prior.  {\bf Below:} Average number of EM iterations vs. the number of sensors $m$.}
\label{fig:solardata}
\end{center}
\end{figure}

Figure \ref{fig:solardata} indicates the performance of the EM MST approximation algorithm by computing the KL divergence between the approximated tree and the original graph. We randomly generate $1000$ different characteristic matrices, $\bH$, and plot the average performance. It is clear in the figure that the proposed EM MST approximated tree achieves better performance comparing to the Chow-Liu optimal tree for the prior graph.
The only problem is over-fitting as it is stated on this figure. The over-fitting occurs, since we are minimizing the KL divergence computed based on the observation vector distribution, $f_{\bS_{\uY}}(\uy)$, instead of the latent vector distribution, $f_{\bSigma}(\ux)$ which is {\it unavailable}. To overcome this, we suggest to learn the maximum number of iterations, $L_{max}$, based on the training data and run the EM algorithm for at most $L_{max}$ iterations. For the set up in this paper we set $L_{max} = 20$.

\section{Conclusion and Future Directions}
\label{sec:conclusion}
We present an iterative algorithm to learn the MST in an underdetermined linear model. The proposed algorithm is a combination of the EM algorithm and the Chow-Liu algorithm. 
We show that the algorithm needs only to run for a limited number of iterations and each iterations can be computed efficiently using the Chow-Liu algorithm.
In future works, we will do a more comprehensive simulation study along with theoretical proof of the EM algorithm convergence. we will also look at the maximum a posteriori solution to the problem as well as learning highly correlated connections in high dimensional data, ``Big Data''.

%\section*{Acknowledgment}
%
%This work was supported in part by NSF grants ECCS-098344, D0E grant, DE-0E0000394, ECCS-1310634, and the
%University of Hawaii REIS project.

\section*{Acknowledgment}

This work was supported in part by NSF grant ECCS-1310634, the Center for Science of Information (CSoI), an NSF Science and Technology Center, under grant agreement CCF-0939370, and the University of Hawaii REIS project.

\bibliographystyle{IEEEbib}
\bibliography{refs}

\begin{thebibliography}{10}

\bibitem{BP}
H.-A. Loeliger, J.~Dauwels, J.~Hu, S.~Korl, L.~Ping, and F.~R. Kschischang,
\newblock ``The factor graph approach to model-based signal processing,''
\newblock {\em Proceedings of the IEEE}, vol. 95, no. 6, pp. 1295--1322, June
  2007.

\bibitem{ChowLiu}
C.~K. Chow and C.~N. Liu,
\newblock ``Approximating discrete probability distributions with dependence
  trees,''
\newblock {\em IEEE Transactions on Information Theory}, vol. 14, no. 3, pp.
  462--467, 1968.

\bibitem{MQ16ArXiv}
N.~T. Khajavi and A.~Kuh,
\newblock ``The quality of the covariance selection through detection problem
  and {AUC} bounds,''
\newblock {\em arXiv preprint arXiv:1605.05776}, 2016.

\bibitem{ITA2016}
N.~T. Khajavi and A.~Kuh,
\newblock ``The quality of tree approximation from auc bounds,''
\newblock {\em Information Theory and Applications Workshop}, 2016.

\bibitem{Allerton16}
N.~T. Khajavi and A.~Kuh,
\newblock ``The goodness of covariance selection problem from {AUC} bounds,''
\newblock in {\em Proc. 54th Annual Allerton Conference on Communication,
  Control, and Computing}, Sep. 2016.

\bibitem{dempster}
A.~P. Dempster,
\newblock ``Covariance selection,''
\newblock {\em Biometrics}, vol. 28, no. 1, pp. 157--175, March 1972.

\bibitem{cover}
T.~M. Cover and J.~A. Thomas,
\newblock {\em Elements of Information Theory},
\newblock Wiley, New York, 1991.

\bibitem{ISIT2013}
N.~T. Khajavi and A.~Kuh,
\newblock ``First order markov chain approximation of microgrid renewable
  generators covariance matrix,''
\newblock in {\em Proc. IEEE International Symposium on Information Theory,
  Istanbul, Turkey (ISIT' 13)}, July 2013, pp. 1207--1211.

\bibitem{NREL}
NREL.~Solar irradiance~website. [Online]. Available:~http://www.nrel.gov/midc/,
  ,'' .

\bibitem{lewis_approx}
P.~M.~Lewis II,
\newblock ``Approximating probability distributions to reduce storage
  requirements,''
\newblock {\em Information and control}, vol. 2, no. 3, pp. 214--225, 1959.

\bibitem{kruskal}
Joseph~B Kruskal,
\newblock ``On the shortest spanning subtree of a graph and the traveling
  salesman problem,''
\newblock {\em Proceedings of the American Mathematical society}, vol. 7, no.
  1, pp. 48--50, 1956.

\bibitem{APSIPA2014}
N.~T. Khajavi, A.~Kuh, and N.~P. Santhanam,
\newblock ``Spatial correlations for solar {PV} generation and its tree
  approximation analysis,''
\newblock in {\em Proceedings of the Asia-Pacific Signal and Information
  Processing Association Annual Summit and Conference (APSIPA ASC)}, Dec 2014,
  pp. 1--5.

\end{thebibliography}

\end{document}